\PassOptionsToPackage{colorlinks=true,linkcolor=blue,citecolor=blue,urlcolor=blue}{hyperref}
\documentclass[preprint]{aastex7}
\usepackage{graphicx}
\usepackage{latexsym,bm}
\usepackage{dcolumn}
\usepackage{orcidlink}
\usepackage{amsmath}
\usepackage{amssymb}
\usepackage{chngcntr}
\usepackage{hyperref}
\usepackage{aas_macros}
\usepackage{threeparttable}
\usepackage{comment}

\begin{document}``

\title{A Two-Population von Mises--Fisher Distribution of the Magnetic Field Direction in the Young Solar Wind}

 \author[orcid=0000-0003-4695-8866,sname='Li']{Gang Li}
\affiliation{State Key Laboratory of Lunar and Planetary Sciences, Macau University of Science and Technology, Macau 999078, China}
\email[show]{gangli.glspl@gmail.com} 

\author[0000-0003-0142-8669]{N. H. Bian}
\affiliation{State Key Laboratory of Lunar and Planetary Sciences and CNSA Macau Center for Space Exploration and Science, 
Macau University of Science and Technology, Macau, China} \email{bnhmaximilien@must.edu.mo}

\author[0000-0002-3595-2525]{S. Y. Huang}
\affiliation{School of Earth and Space Science and Technology, Wuhan University, Wuhan, 430072, China} 
\email{shiyonghuang@whu.edu.cn}

\author[0000-0002-8932-0793]{A. Chian}
\affiliation{School of Mathematical Sciences, Adelaide University, Adelaide, SA5005, Australia}
\affiliation{National Institute for Space Research (INPE), Sao Jose dos Campos, SP 12 227010, Brazil}
\email{abraham.chian@gmail.com}

\author[0000-0001-6880-4468]{R. Santos-Lima}
\affiliation{Instituto de Astronomia, Geof\'isica e Ci\^encias Atmosf\'ericas, Universidade de S\~ao Paulo (IAG-USP), R. do Mat\~ao 1226, S\~ao Paulo, SP 05508-090, Brazil}
\email{reinaldo.lima@iag.usp.br}

\author[0000-0002-4971-5854]{E. L. Rempel}
\affiliation{Aeronautics Institute of Technology (ITA), S\~ao Jos\'e dos Campos, SP, 12228-900, Brazil}
\email{rempel@yahoo.com.br}

\author[0000-0002-9861-0557]{R. A. Miranda}
\affiliation{Faculty of Science and Technology in Engineering, and Plasma Physics Laboratory, Institute of Physics, University of Bras\'ilia (UnB), Bras\'ilia DF 70910-900, Brazil}
\email{rmiracer@unb.br}

\author[0000-0002-0744-9746]{B. Raphaldini}
\affiliation{Instituto de Matemática, Estatística e Ci\^encia da Computaç\~ao da, Universidade de S\~ao Paulo (IAG-USP), S\~ao Paulo, SP 05508-090, Brazil}
\email{}

\author[0000-0003-4682-2459]{Yasmmin Tamburus}
\affiliation{Instituto de Astronomia, Geof\'isica e Ci\^encias Atmosf\'ericas, Universidade de S\~ao Paulo (IAG-USP), R. do Mat\~ao 1226, S\~ao Paulo, SP 05508-090, Brazil}
\email{ytamburus@usp.br}


\date{\today}

\begin{abstract}
The one-point distribution of the magnetic field direction in the young solar wind is
found to be a superposition of two populations rather than a single exponential. Using
Parker Solar Probe (PSP) Encounter~1 magnetometer data, we show that the distribution of
the pitch-angle cosine $\mu(t) = \hat{\mathbf b}(t)\cdot\hat{\mathbf b}_0$ is well
described by a two-component von Mises--Fisher (vMF) mixture: a broad, weakly confined
component of concentration $\kappa_1 \simeq 3.6$ that accounts for the large-amplitude
rotations and switchback tail, and a field-aligned component that reduces, in the
small-angle limit, to a Gaussian in the polar angle $\theta$ of width
$\sigma \simeq 15^\circ$ (concentration $\kappa_2 \simeq 14$). The concentration of the
field-aligned component is not a free parameter: it is set by the level of the
transverse Alfv\'enic fluctuation through
$\kappa_2 = 2\langle |\mathbf B|^2\rangle/\langle \delta B_\perp^2\rangle
= 2/\langle 1-\mu^2\rangle \simeq 16.2$, in agreement with the fitted value to within
${\sim}12\%$. The transverse fluctuations are weakly anisotropic
($\langle\delta B_1^2\rangle/\langle\delta B_2^2\rangle \simeq 1.1$), so the core is a
weakly anisotropic Bingham distribution rather than a pure vMF. The broad component is
exponential in $\mu$, not in $\theta$, confirming that switchbacks saturate toward
$\mu = -1$ rather than decaying away, consistent with the drift--diffusion framework of
\citet{Bian2024ApJ...960L...15L}.  This physics-based two-population framework connects macroscopic fluid parameters directly to those describing the directional statistics, offering a powerful diagnostic for probing the structural evolution of MHD turbulence and switchbacks in the inner heliosphere.
\end{abstract}


\section{Introduction}

The magnetic field vector $\mathbf B(t)$ measured by a spacecraft in the turbulent solar
wind deflects persistently from the mean field direction $\hat{\mathbf b}_0$ \citep{Parker1958ApJ...128...664P,Jokipii1966ApJ...146...480J}. 
Such fluctuations have been examined in the context of extension of  coronal field into the heliosphere by \citet{Ness1966ApJ...143...23N, Ness1967SoPh....2..351N}.  At 1 au,  the distribution of the angle between the instantaneous $\mathbf B(t)$ and the average field,  over different solar wind conditions, have been examined in, e.g.,  \citep{Borovsky2010JGR...115...A09101B, Xu2015ApJ...801...58L, Chang2022ApJ...931...105H}.  
The entire spectrum of variations from the Parker field can be attributed to dynamical processes occurring in the plasma. In Alfv\'enic streams, the field magnitude remains close
to a constant, so that the evolution of $\mathbf B(t)$ can be understood from that of 
the tip of the unit vector $\hat{\mathbf b}(t) = \mathbf B(t)/|\mathbf B(t)|$, which is confined to the unit sphere. 
The one-point statistics of   $\hat{\mathbf b}(t)$ is conveniently described by the pitch-angle cosine,
\begin{equation}
  \mu(t) = \hat{\mathbf b}(t)\cdot\hat{\mathbf b}_0 = \cos\theta(t),
  \label{eq:mu}
\end{equation}
where $\theta$ is the polar angle measured from the mean field direction. The random walk of
$\hat{\mathbf b}$ on the unit sphere 
was first treated by
\citet{Barnes1981JGR...86...7498B} as a pure rotational diffusion, which relaxes the field
direction toward an isotropic distribution with no preferred direction.
\citet{Bian2024ApJ...960L...15L} generalized this to a steady-state advection--diffusion
(drift--diffusion) model, in which a mean-field restoring force advects $\hat{\mathbf b}$ back
toward $\hat{\mathbf b}_0$ against the stochastic diffusion; the balance of the two yields the
exponential (equivalently, von Mises--Fisher, vMF) distribution
\begin{equation}
  P(\mu) = \frac{\kappa}{2\sinh\kappa}\,\exp(\kappa\mu),
  \qquad \mu \in [-1,1],
  \label{eq:vMF}
\end{equation}
with concentration parameter $\kappa$ related to the mean pitch-angle cosine by the
Langevin function $\langle\mu\rangle = \coth\kappa - 1/\kappa$. The exponential form was
established observationally by \citet{Bavassano1983}, \citet{Borovsky2008JGR...13...A08110B}, and
\citet{DudokdeWit2020ApJS...246...39W}; for large $\kappa$ the exponential in $\mu$ reduces to a Gaussian in
$\theta$, the standard fitting practice to the angular data at small angles.

A single exponential, however, provides only a first description. Two distinct features of the young solar wind suggest that a one-parameter fit is inadequate. First, the magnetic field direction
exhibits a broad, nearly isotropic component associated with large-amplitude rotations and
reversals of the radial field (switchbacks; \citealt{Bale2019Nat...576...237W};
\citealt{Fargette2021ApJ...919...96B,Fargette2022A&A...663...A109K}), which populate the tail $\mu \rightarrow -1$. Second,
the bulk of the fluctuations are small-angle, nearly transverse deflections about the mean field \citep{Belcher1971JGR...76...3534D}, which produce a concentrated excess near $\mu = 1$. A natural and minimal
description is therefore a \emph{two-population} directional distribution. Two-population
fits to the angular deflection of the magnetic field has been practiced by \citealt{Xu2015ApJ...801...58L}, who fitted the angular distribution by the sum of a Gaussian core and an exponential tail {in angle} (instead of $\mu$.).

In this Letter we carry out a two-population analysis of the PSP magnetic field data of Alfv\'{e}nic streams within
the directional-statistics framework of \citet{Bian2024ApJ...960L...15L}. We show that $P(\mu)$ has two vMF populations, one broad and one field-aligned. We also show that the concentration parameter of the field-aligned component vMF is \emph{regulated}, without any free parameter, by the transverse Alfv\'enic fluctuation level. 
We clarify that the functional form of the broad component as exponential in $\mu$, not in $\theta$ as in \citep{Xu2015ApJ...801...58L}.

\section{Data and Method}

\begin{figure*}[t]
  \centering
  \includegraphics[width=0.8\textwidth]{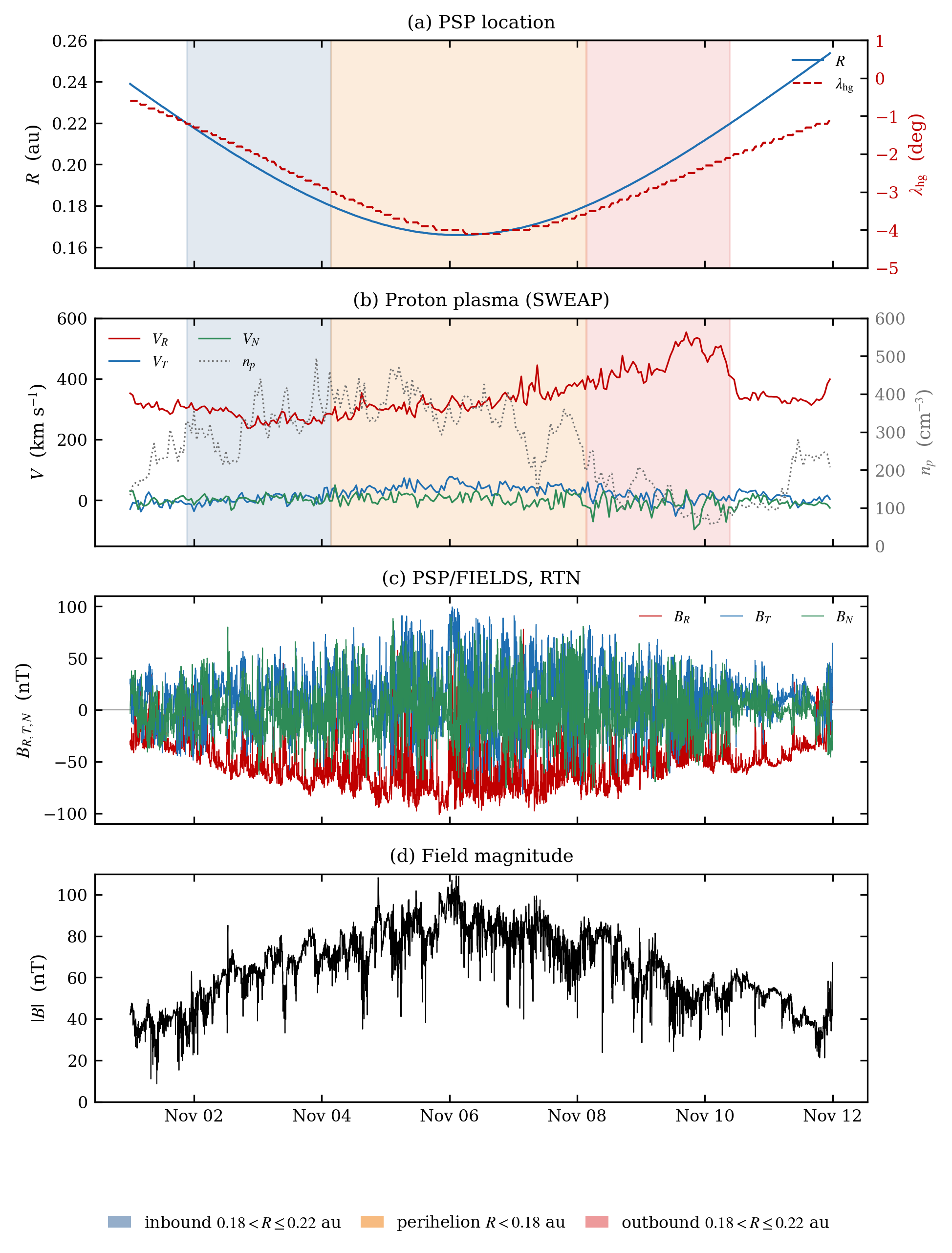}
  \caption{PSP Encounter~1 (2018 November 1--11) overview. (a) Heliocentric distance
  $R$ (left axis) and heliographic latitude $\lambda_{\rm hg}$ (right axis).
  (b) Proton plasma from SWEAP: velocity components $V_{R,T,N}$ (left axis) and number
  density $n_p$ (right axis, dotted). (c) The three RTN magnetic-field components and
  (d) the field magnitude $|\mathbf B|$, from PSP/FIELDS at 1~minute resolution. The
  shaded bands in (a) and (b) mark the three radial ranges: inbound $0.18<R\le0.22$~au,
  perihelion $R<0.18$~au, and outbound $0.18<R\le0.22$~au.}
  \label{fig:Fig1_context}
\end{figure*}

We use the PSP/FIELDS fluxgate magnetometer \citep{Bale2016SSR...204...49W}, Level-2 RTN vectors at 1~minute resolution (PSP\_FLD\_L2\_MAG\_RTN\_1MIN), over Encounter~1 (2018 November 1--11), during which the heliocentric distance $R$ decreases from ${\sim}0.25$ au to $0.17$ au at perihelion and then returns to ${\sim}0.25$ au. After discarding $22$ NaNs, the series contains $15{,}818$ data points. 
We separate the encounter into three intervals in heliocentric distance
(Figure~\ref{fig:Fig1_context}): the inbound leg $0.18<R\le0.22$~au, the perihelion region
$R<0.18$~au, and the outbound leg $0.18<R\le0.22$~au. Within each interval the background field
$\mathbf B_0(t)$ is a 6~hr moving average of the full vector field, and the pitch-angle cosine is
$\mu(t) = \hat{\mathbf b}(t)\cdot\hat{\mathbf b}_0(t)$ with
$\hat{\mathbf b}_0(t) = \mathbf B_0(t)/|\mathbf B_0(t)|$; the mean pitch-angle cosine is
$\langle\mu\rangle = 0.81$.

Figure~\ref{fig:Fig1_context} presents the macroscopic plasma and magnetic field conditions during Encounter 1. As the spacecraft descends from $0.25$~au to a perihelion of $\sim 0.17$~au and back out, it remains immersed in a highly Alfvénic, relatively slow solar wind stream ($V_R \sim 300\text{--}400 \mathrm{~km~s^{-1}}$). The magnetic field is dominated by a sunward (negative) radial polarity. While the individual RTN magnetic field components (Panel c) exhibit intense, high-frequency fluctuations and frequent directional reversals (switchbacks), the total magnetic field magnitude $|\mathbf{B}|$ (Panel d) evolves smoothly with radial distance and exhibits comparatively little high-frequency variance. This distinct contrast---violent directional scattering alongside a stable field magnitude---confirms that the fluctuations are predominantly transverse and spherically polarized.

We model the normalized distribution of $\mu$ by a mixture of two vMF (exponential)
distributions,
\begin{equation}
  P(\mu) = w\,\frac{\kappa_1}{2\sinh\kappa_1}\exp(\kappa_1\mu)
         + (1-w)\,\frac{\kappa_2}{2\sinh\kappa_2}\exp(\kappa_2\mu),
  \label{eq:mixture}
\end{equation}
where $w$ is the mass fraction of the broad component. For $\kappa_2 \gg 1$ the second
term reduces, using $\cos\theta \simeq 1 - \theta^2/2$, to a Gaussian in the polar angle,
\begin{equation}
  P_2(\mu) \propto \exp\!\left(-\frac{\theta^2}{2\sigma^2}\right),
  \qquad \theta = \arccos\mu,
  \qquad \sigma = \kappa_2^{-1/2}.
  \label{eq:gaussian}
\end{equation}


A total of $40$ bins in $\mu$ from $-1.00$ to $1.00$, with a bin width of $0.05$ were used in binning the data.  We minimize 
the Baker-Cousins modified Poisson likelihood ratio $\chi_{BC}^2$
between the model and the binned histogram of $\mu$.  
Here $\chi^2_{BC}$ is defined by \citep{Baker1984},
\begin{equation}
    \chi^2_{\mathrm{BC}} = 2 \sum_{i=1}^{N_{\mathrm{bins}}} \left[  M_i(\mathbf{\kappa_1, \kappa_2}, w) - N_i + N_i \ln\left(\frac{N_i}{ M_i(\mathbf{\kappa_1, \kappa_2}, w)} \right )\right],
\end{equation}
where $N_i$ is the measured count in the $i$-th angular bin, $M_i(\mathbf{\kappa_1, \sigma}, w)$ is the expected model count, and $N_{\mathrm{bins}} = 40$ is the total number of equally spaced bins spanning $\mu \in [-1, 1]$.

To restrict the directional statistics to Alfv\'enic streams, we compute the normalized
cross helicity and the residual energy from the SWEAP proton velocity $\mathbf v$ and the
Alfv\'en-normalized magnetic field $\mathbf b = \mathbf B/\sqrt{\mu_0\rho}$,
\begin{equation}
  \sigma_c = \frac{2\langle\delta\mathbf v\cdot\delta\mathbf b\rangle}
                    {\langle|\delta\mathbf v|^2\rangle+\langle|\delta\mathbf b|^2\rangle},
  \qquad
  \sigma_r = \frac{\langle|\delta\mathbf v|^2\rangle-\langle|\delta\mathbf b|^2\rangle}
                    {\langle|\delta\mathbf v|^2\rangle+\langle|\delta\mathbf b|^2\rangle},
  \label{eq:sigc}
\end{equation}
where $\delta\mathbf v$ and $\delta\mathbf b$ are the fluctuations about the local mean
within a window. Because the proton velocity is reliable only at 1~hour resolution in the
COHOWeb merged product, $\sigma_c$ and $\sigma_r$ are evaluated over 6~hr windows at 1~hr
cadence (full RTN vector).  A window is deemed to be Alfv\'enic if $\sigma_c > 0.5$ and we find that 
 $212$ of $258$ windows (82\%) are Alfv\'enic.
The background field $\mathbf B_0(t)$ used to form
$\mu(t) = \hat{\mathbf b}(t)\cdot\hat{\mathbf b}_0(t)$ is a 6~hr boxcar average of the full
vector field, with the data point at the center of the window. A one-minute $\mu$ point is
retained (``tallied'') only if more than half of its 6~hr window is Alfv\'enic. As a result, 12,914 of 15,430 points (84\%) survive this selection criterion.

\begin{figure}[t]
  \centering
  \includegraphics[width=\textwidth]{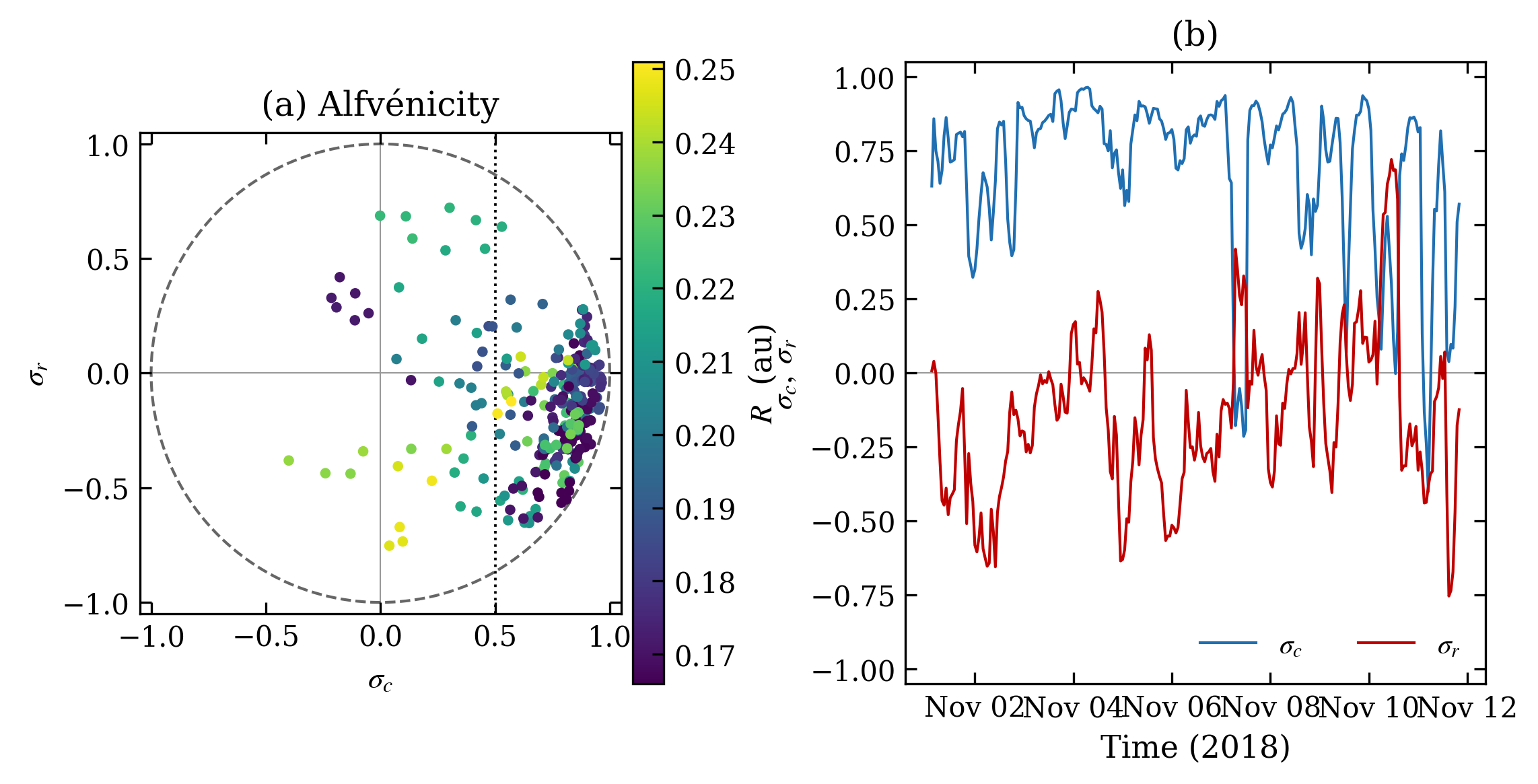}
  \caption{Alfv\'enicity of PSP Encounter~1 from 6~hr windows at 1~hr cadence (full RTN
  vector; $\mathbf b = \mathbf B/\sqrt{\mu_0\rho}$).
  (a) Residual energy $\sigma_r$ versus normalized cross helicity $\sigma_c$, colored by
  heliocentric distance $R$, with the unit circle $\sigma_c^2+\sigma_r^2=1$ (dashed) and the
  Alfv\'enicity threshold $\sigma_c=0.5$ (dotted); most points cluster near the circle,
  confirming that this interval is strongly Alfv\'enic. (b) $\sigma_c$ (blue) and $\sigma_r$
  (red) versus time.}
  \label{fig:Fig2_alfvenicity}
\end{figure}

Figure~\ref{fig:Fig2_alfvenicity} shows $\sigma_r$ versus $\sigma_c$ (colored by
heliocentric distance) and their time series. The interval is strongly Alfv\'enic: the
fluctuations lie near the $\sigma_c^2+\sigma_r^2=1$ circle, with mean $\sigma_c \simeq 0.69$
and mean $\sigma_r \simeq -0.14$, i.e., near kinetic--magnetic equipartition with a weak
magnetic excess. Binning the tallied $\mu$ points into the three radial ranges
(Figure~\ref{fig:Fig1_context}) gives 2995, 5332, and 2523 points (92\%, 93\%, and 77\% of each range), and the two-component fit shows a weak radial trend as we discuss in the next section. 


\section{Results and Discussion}

\begin{figure}[t]
  \centering
  \includegraphics[width=0.8\textwidth]{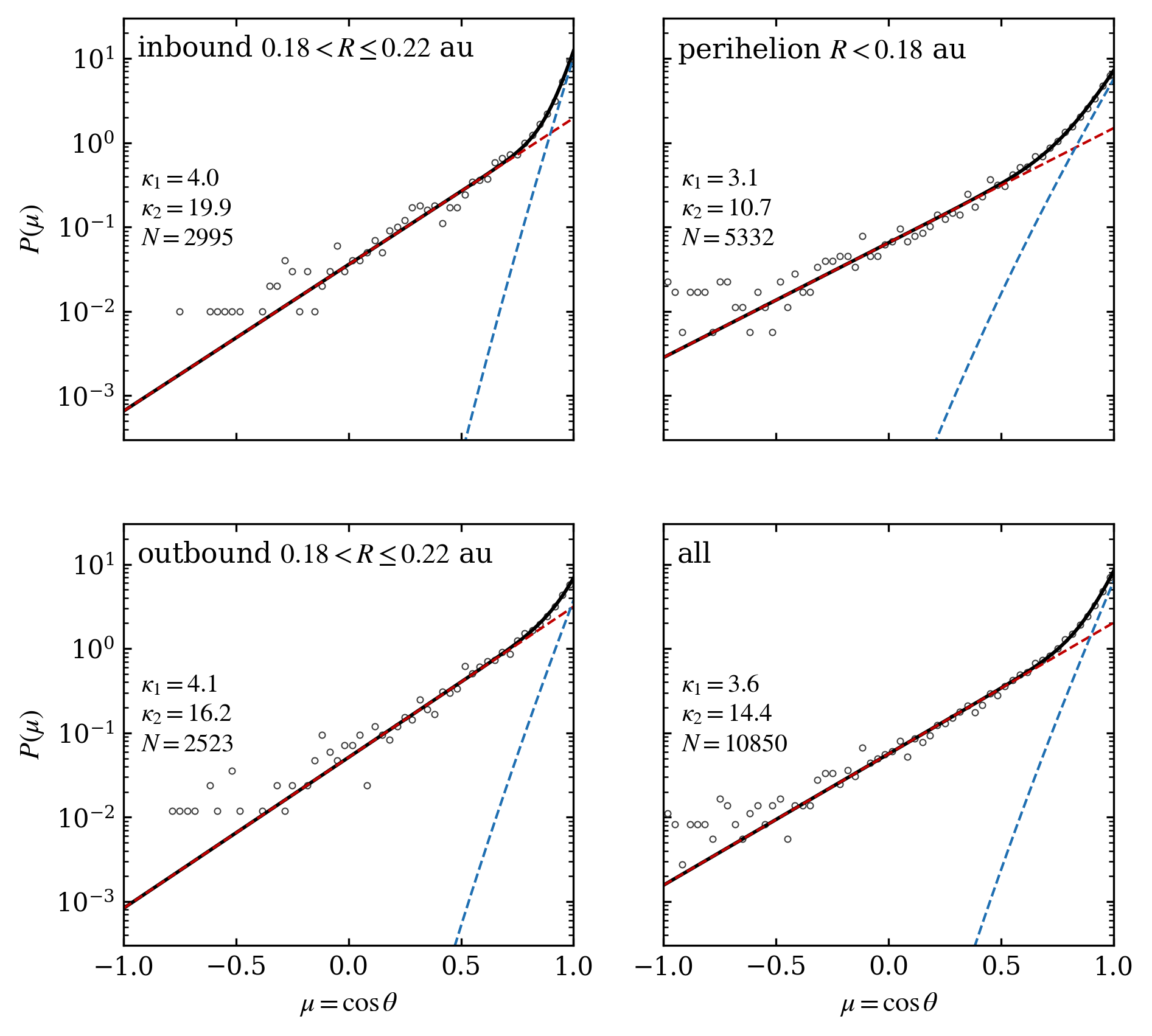}
  \caption{Two-component fit to the pitch-angle cosine $\mu$ (log--linear) for the
  Alfv\'enic periods of PSP Encounter~1, in three ranges of heliocentric distance $R$
  (inbound $0.18<R\le0.22$~au, perihelion $R<0.18$~au, outbound $0.18<R\le0.22$~au) and,
  in the bottom-right panel, all three combined. In each panel the broad exponential tail
  (red dashed line) dominates the switchback tail $\mu\rightarrow-1$ and the Gaussian-in-angle core
  (blue dashed line) dominates near $\mu=1$; the fitted $\kappa_1$, $\kappa_2$,  and the number of points $N$ are given in each panel.}
  \label{fig:Fig3_fitting}
\end{figure}

Figure~\ref{fig:Fig3_fitting} shows the binned distribution of $\mu$ together with the
two-component fit, in three ranges of heliocentric distance separating the inbound and
outbound legs of the encounter, plus the combined sample.
By separating the tallied data into the three heliocentric radial ranges, we observe a distinct spatial evolution in the two-population structure, reflecting the radial development of the turbulence (see Figure~\ref{fig:Fig3_fitting}). Approaching the Sun, the core concentration drops significantly from the inbound leg ($\kappa_2 \simeq 20$) to the perihelion region ($\kappa_2 \simeq 11$), indicating a broadening of the field-aligned core as the relative transverse fluctuation level $\langle \delta B_\perp^2 \rangle / \langle |\mathbf{B}|^2 \rangle$ intensifies deeper in the corona. Concurrently, the broad tail component also widens at perihelion ($\kappa_1 \simeq 3.1$) compared to the inbound leg ($\kappa_1 \simeq 4.0$), corresponding to an increased occurrence rate and angular spread of switchbacks. The outbound leg exhibits intermediate concentrations ($\kappa_2 \simeq 16$, $\kappa_1 \simeq 4.1$), which may reflect intrinsic longitudinal asymmetries or different solar wind source streams crossed by the spacecraft during its trajectory. Despite this radial variation, the two-component vMF mixture accurately captures the distribution across all distances, confirming its universality in the young solar wind.

On the combined data, the broad component has concentration $\kappa_1 \simeq 3.6$ and fraction $w \simeq 0.57$ and the field-aligned core component has $\kappa_2 \simeq 14.4$ (equivalent to a width of $\sigma \simeq 15.1^\circ$) and a fraction of $1-w \simeq 0.43$.
The two-component fit gives $R^2 = 0.999$ over the full range $[-1,1]$.
The broader component of this mixture ($\kappa_1 \simeq 3.6$) include in it the ubiquitous magnetic field switchbacks observed extensively in the inner heliosphere. These structures represent large-amplitude, nonlinear Alfv\'{e}nic fluctuations evolving on top of the background field. Large-amplitude spherically polarized Alfv\'{e}n waves have been established in earlier foundational work by \citet{Barnes1976}. As the exact nonlinear solutions to the ideal magnetohydrodynamic equations, these are fluctuations possessing a strictly conserved magnetic field magnitude ($|\mathbf{B}| = \text{constant}$).
Later, \citet{Barnes1981JGR...86...7498B} developed a statistical theory for Alfv\'{e}nic disturbances, in which the magnetic field moves randomly over the surface of a sphere of radius $|B|$. 
When incorporating these insights into modern interpretations of Parker Solar Probe observations, the broader tail component ($\kappa_1$) of the angular distribution naturally captures the large-amplitude arc-polarized nature of these waves. 
Recently  \citet{Bian2024ApJ...960L...15L} showed that these structures occur naturally as the steady state solution of an advection-diffusion process of the tip of the B field on a sphere, extending the work by \citet{Barnes1981JGR...86...7498B}.

Comparing to the broad component, the core component is fully governed by the small-amplitude transverse Alfv\'{e}nic fluctuations, and $\kappa_2$ can be related to 
small-amplitude transverse Alfv\'{e}nic fluctuation level.  This can be seen from the following.  Let the unit vectors
$\mathbf e_1, \mathbf e_2$ span the plane perpendicular to the mean direction
$\hat{\mathbf b}_0$, with $\mathbf e_1$ in the radial--tangential (RT) plane and
$\mathbf e_2 \simeq -\hat{\mathbf n}$ along the normal. The transverse deflection of the
${\mathbf B}$ field is then, 
\begin{equation}
  \delta {\mathbf B}_\perp = \delta B_1\,\mathbf e_1 + \delta B_2\,\mathbf e_2,
  \qquad
  \delta B_1 = \delta {\mathbf b}_\perp \cdot\mathbf e_1,\quad
  \delta B_2 = \delta {\mathbf b}_\perp \cdot\mathbf e_2,
  \label{eq:transverse}
\end{equation}
with $|\delta {\mathbf B}_\perp|^2 = \delta B_1^2 + \delta B_2^2 = (1-\mu^2) B_0^2$. Note that $\mu$ is the cosine of the angle between ${\mathbf B}$ and ${\mathbf B_0}$, and spherical polarization still holds. If we model the small-amplitude Alfv\'enic fluctuation as two \emph{independent} Gaussian random variables, along the two transverse directions and with root-mean-square $(\sigma_{1,2} B_0)^2$ ($\sigma_{1,2}$ are dimensionless), then we have
\begin{equation}
  P(\delta b_1, \delta b_2)\,
    d\delta b_1\, d\delta b_2
  = \frac{1}{2\pi\sigma_1\sigma_2  B_0^2}\,
    \exp\!\left[-\frac{(\delta b_1/B_0)^2}{2\sigma_1^2}
                    -\frac{(\delta b_2/B_0)^2}{2\sigma_2^2}\right]
    d\delta b_1\, d\delta b_2.
  \label{eq:gauss2d}
\end{equation}
Note that the tangent-plane area element coincides with the solid-angle element at small angles ($d\delta b_1\,d\delta b_2 \simeq B_0^2 \sin\theta\,d\theta\,d\phi$).
Eq.~\eqref{eq:gauss2d} is the \emph{Bingham} distribution on the unit sphere when $\sigma_1 \neq \sigma_2$.

Assuming $\sigma_1 = \sigma_2 = \sigma_\perp$, 
the Bingham distribution collapses to the vMF distribution,  
\begin{equation}
  P(\delta {\mathbf b}) \propto
    \exp\!\left[-\frac{1-\mu^2}{2\sigma_\perp^2}\right]
    \simeq \exp(\kappa\mu),
  \qquad
  \kappa = \frac{1}{\sigma_\perp^2},
  \label{eq:kappa_unit}
\end{equation}
where $\mu^2-1 = (\mu-1)(\mu+1) \simeq 2(1-\mu)$ at small angles. 
Because $\langle 1-\mu^2\rangle = (\langle\delta b_1^2\rangle + \langle\delta b_2^2\rangle)/B_0^2 = \sigma_1^2 + \sigma_2^2$ (equal to $2\sigma_\perp^2$ in the
isotropic case), therefore the core component is 
governed by the transverse fluctuation level,
\begin{equation}
  \kappa_2 = \frac{2}{\langle 1-\mu^2\rangle} \approx \frac{2\,\langle|\mathbf B|^2\rangle}{\langle\delta B_\perp^2\rangle}.
  \label{eq:kappa_circ}
\end{equation}
In the last step, we make use of the fact that the field magnitude is nearly constant for Alfv\'enic fluctuations, and compute it using $\sqrt{\langle|\mathbf B|^2 \rangle}$.

As a consistency check, the posterior-weighted average $\langle 1-\mu^2\rangle $ 
of the core is $0.124$ for the entire encounter, yielding
$\kappa_2 = 2/\langle 1-\mu^2\rangle = 16.2$. Alternatively, the 
posterior-weighted average $\langle\delta B_\perp^2\rangle = 670$~nT$^2$ and $\langle|\mathbf B|^2\rangle = 5491$~nT$^2$, giving a  relative transverse fluctuation level 
$\langle\delta B_\perp^2\rangle/\langle|\mathbf B|^2\rangle = 0.122$, hence $\kappa_2 = 2\langle|\mathbf B|^2\rangle/\langle\delta B_\perp^2\rangle = 16.4$ from the second part of equation~\eqref{eq:kappa_circ}.  Both agree with the fitted $\kappa_2 = 14.4$ to within ${\sim}12\%$, and this agreement holds in each radial range, the fitted $\kappa_2$ vs that using $2/<1-\mu^2>$ are: ($19.9$ vs $21.8$ for inbound, $10.7$ vs $12.5$ for perihelion,  $16.2$ vs $17.9$ for outbound; see Figure~\ref{fig:Fig3_fitting}).  We remark that the posterior weights used in obtaining $<1- \mu^2>$ are drawn from the two-component fit, therefore the above calculation is not fully independent of the fit. Nevertheless, the agreement of $\kappa_2$ between equation~\eqref{eq:kappa_circ} and the fitting provides a meaningful consistency check, as it indicates that the core concentration follows from the transverse-fluctuation moments. In part, the $\sim 10\%$ discrepancy between the estimation and the fitting of  $\kappa_2$ is due to the fact that there is a weak transverse anisotropy between $\delta B_{e_1}^2$ and 
$\delta B_{e_2}^2$, so that the exact core is a Bingham rather than a vMF. 

The emergence of a vMF distribution is explained  using the  Fokker-Planck framework by \citet{Bian2024ApJ...960L...15L}, where 
a drift diffusion process with a constant restoring rate is assumed, yielding the one-population exponential of Eq.~\eqref{eq:vMF}. The observed two-population structure indicates that a constant restoring rate is insufficient.  
A natural extension of \citet{Bian2024ApJ...960L...15L} is to consider a $\mu$-dependent restoring rate $\Omega(\mu)$ which is larger near $\mu = 1$ (strongly focusing the small-angle core) and decreases as  $\mu$ decreases, accommodating the observed switch-back populations. 

\begin{figure}[t]
  \centering
  \includegraphics[width=0.9\textwidth]{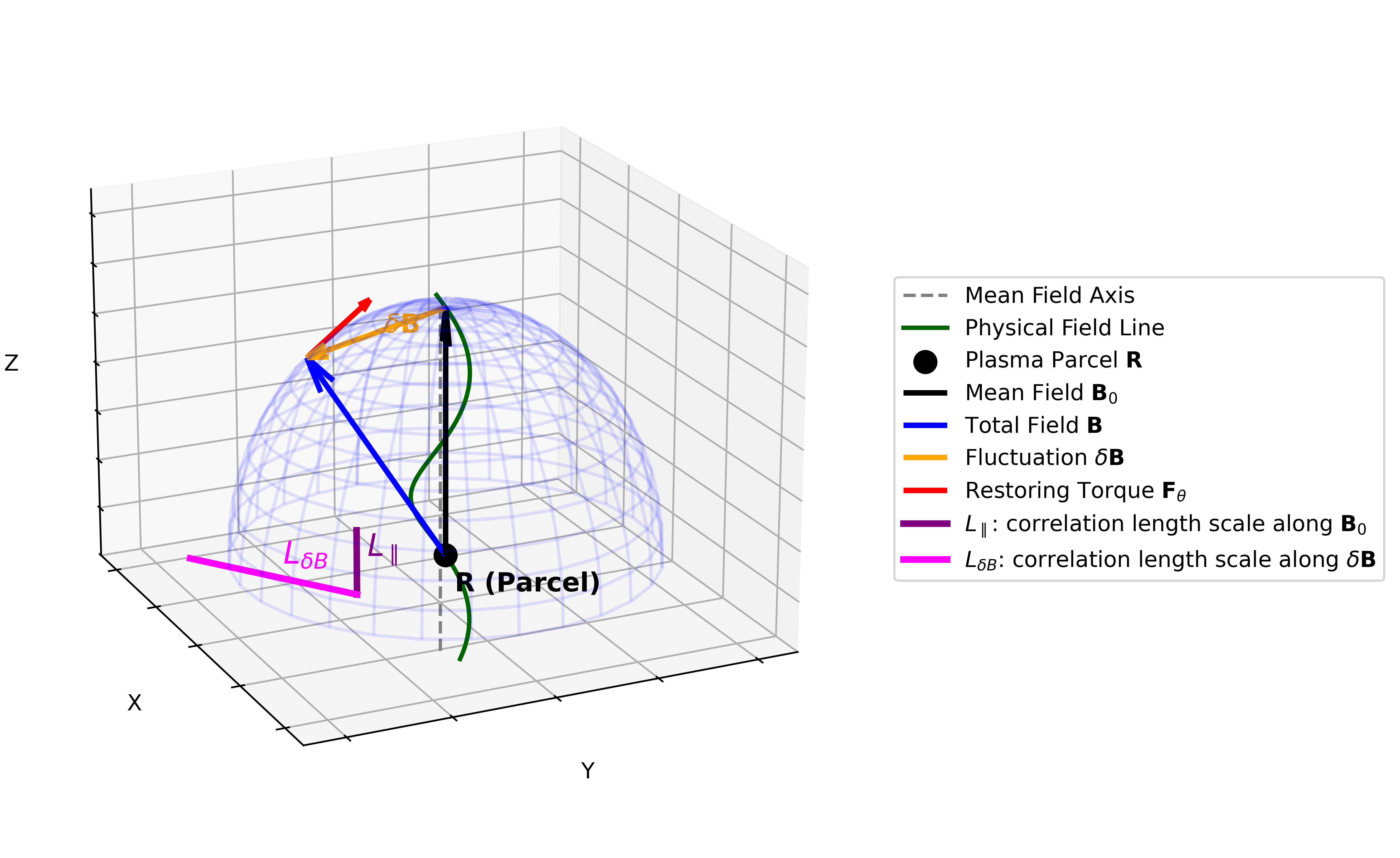}
  \caption{
  3D geometric visualization of a plasma parcel (black dot) embedded in a spherically polarized Alfvénic wave packet. The physical field line (green) restricts the local magnetic field $\mathbf{B}$ (blue) and its perturbation $\delta\mathbf{B}$ (orange) to a constant-$\vert{}\mathbf{B}\vert{}$ sphere. The macroscopic restoring torque $\mathbf{F}_\theta$ (red) drives the relaxation of the pitch angle $\theta$. This torque is composed of two competing spatial gradients: the linear background tension acting over the parallel correlation length $L_\parallel$ (purple), and the non-linear self-tension acting over the active perturbation scale $L_{\delta B}$ (magenta). As detailed in the text, the dynamic competition between these two length scales determines the state-dependent relaxation rate $\Omega(\theta)$.}
  \label{fig:Fig4_SAW}
\end{figure}

The geometric and physical basis for this state-dependent restoring force is visualized in Figure 4, which illustrates a plasma parcel embedded in a spherically polarized Alfvénic wave packet. Because the plasma is tied to the physical field line, its transverse velocity fluctuation $\delta\mathbf{v}$ is parallel to the magnetic perturbation $\delta\mathbf{B}$ and strictly perpendicular to the total local field $\mathbf{B}$. In the constant-$\vert{}\mathbf{B}\vert{}$ spherical phase space, this transverse motion naturally decomposes into azimuthal ($\hat{\phi}$) and polar ($\hat{\theta}$) directions. The azimuthal component represents phase-mixing and does not contribute to the 
restoring torque. The polar restoring torque $F_\theta$, however, drives the relaxation of the pitch angle $\theta$ and is composed of two distinct spatial gradients. As demonstrated below, the linear background tension dominates at small deflection angles (acting over the parallel correlation length $L_\parallel$), while the nonlinear self-tension becomes dynamically significant at large deflection angles (acting over a length scale along $\delta B$,  which we denote as $L_{\delta B}$). This physical decomposition reveals why the macroscopic relaxation rate $\Omega$ is fundamentally $\theta$-dependent, providing the origin for the two distinct populations in the observed angular distribution.

We now show that a $\mu$-dependent restoring force can be understood by
considering a fluid parcel's motion in Alfv\'enic fluctuations. 
Let the parcel have a mass $m$. The parcel's motion is governed by Newton's second law, driven by three primary forces: a macroscopic restoring force from magnetic tension $\mathbf{F}_T$, a macroscopic turbulent drag or damping from the surrounding plasma, $-\gamma_p {\delta\mathbf{v}_\perp}$ (see \citep{Li2023ApJ...945...150B}), and a rapidly fluctuating stochastic force representing the random impacts of turbulent eddies ($\boldsymbol{\Gamma}(t)$):
\begin{equation}
    m \frac{d\mathbf{ \delta v}_\perp}{dt} = \mathbf{F}_T - \gamma_p {\delta \mathbf{v}_\perp} + \boldsymbol{\Gamma}(t).
\end{equation}
For Alfv\'enic fluctuations, the parcel velocity and the magnetic perturbation are connected by the Wal\'en relation:
$    \delta\mathbf{v}_\perp = \pm \frac{\delta\mathbf{B}_\perp}{\sqrt{\mu_0 \rho}}$,  and due to the constant-$|\mathbf{B}|$ spherical polarization constraint, the field vector is confined to the surface of a sphere. We can therefore uniquely parameterize both the displacement of the plasma parcel and the deflection of the magnetic field vector by the polar angle $\theta$ relative to the mean field $\mathbf{B}_0$. The transverse velocity (with respect to the total $\mathbf{B}$ field) is directly proportional to the angular drift ($\delta v_\perp \propto d\theta/dt$).  Furthermore if we assume the drag from turbulent scattering instantly overwhelms the parcel's inertia, then the acceleration term vanishes and we have,
\begin{equation}
    \mathbf{\delta v}_\perp = \frac{1}{\gamma_p} \mathbf{F}_T + \frac{1}{\gamma_p} \boldsymbol{\Gamma}(t).
    \label{eq:overdamped_v}
\end{equation}

Because the transverse velocity fluctuation $\delta\mathbf{v}_\perp$ and the magnetic perturbation $\delta\mathbf{B}$ are parallel, the angular coordinate $\theta$ tracking the fluid parcel's position is geometrically identical to the pitch angle between the total field $\mathbf{B}$ and the mean field $\mathbf{B}_0$. The generalized angular force $F_\theta$ driving the Langevin equation is simply the polar projection of the macroscopic magnetic tension force density, $F_T = \frac{1}{\mu_0}(\mathbf{B} \cdot \nabla)\mathbf{B}$, onto the tangent of the constant-$|\mathbf{B}|$ sphere.

Decomposing the field into a background and a fluctuation ($\mathbf{B} = \mathbf{B}_0 + \delta\mathbf{B}$), the tension force splits into a linear and a non-linear component:
\begin{equation}
F_T = \frac{1}{\mu_0}(\mathbf{B}_0 \cdot \nabla)\delta\mathbf{B} + \frac{1}{\mu_0}(\delta\mathbf{B} \cdot \nabla)\delta\mathbf{B}.
\label{eq:tension_split}
\end{equation}
To evaluate these spatial gradients, we follow standard phenomenological scaling for MHD turbulence \citep{Goldreich1995ApJ...438...763S}, and assume the two terms operate over fundamentally different correlation length scales (Figure~\ref{fig:Fig4_SAW}). 

The linear term represents the spatial derivative strictly along the mean field $\mathbf{B}_0$. Approximating this gradient over the parallel correlation length $L_\parallel$, the background tension acts as a restoring force pulling the perturbation straight, yielding a 3D force $\mathbf{F}_{\text{lin}} \simeq -(\mu_0 L_\parallel)^{-1} B_0 \delta\mathbf{B}$. Projecting this physical pull onto the spherical tangent vector $\hat{\theta}$ isolates the effective macroscopic restoring torque, which is strictly proportional to the sine of the deflection angle:

\begin{equation}
F_{\theta,\text{lin}} = \mathbf{F}_{\text{lin}} \cdot \hat{\theta} \simeq -\frac{B_0^2}{\mu_0 L_\parallel} \sin\theta. \label{eq:F_linear}
\end{equation}

Conversely, the non-linear self-tension term represents the gradient along the fluctuation $\delta\mathbf{B}$ itself. In turbulence phenomenology, variations along this active perturbation occur over its own correlation scale, $L_{\delta B}$ (which replaces the strictly transverse eddy scale $L_\perp$ because $\delta\mathbf{B}$ is not perpendicular to $\mathbf{B}_0$ at large amplitudes), giving the scalar operator $(\delta\mathbf{B} \cdot \nabla) \simeq \vert{}\delta\mathbf{B}\vert{}/ \mathbf{L_{\delta B}}$. Because spherical polarization rigorously constrains the fluctuation magnitude to $\vert{}\delta\mathbf{B}\vert{}^2 = 2B_0^2(1-\cos\theta)$, applying this operator yields a non-linear force magnitude $\vert{}\mathbf{F}_{\text{nl}}\vert{} \simeq (2B_0^2 / \mu_0 \mathbf{L_{\delta B}})(1-\cos\theta)$. When projected onto the meridian, this self-tension acts to inflate the folded field line, providing a counter-torque that opposes the linear background tension:

\begin{equation}
F_{\theta,\text{nl}} = \mathbf{F}_{\text{nl}} \cdot \hat{\theta} \simeq +\frac{2B_0^2}{\mu_0 {L_{\delta B}}} (1-\cos\theta) \sin\theta.
    \label{eq:F_nl}
    \end{equation}

Modeling the turbulent impacts $\Gamma(t)$ as Gaussian white noise requires care, as the fluctuations are constrained to the constant-$|\mathbf{B}|$ spherical surface. For Brownian motion on a Riemannian manifold with metric $g_{ij}$, the It\^{o} stochastic differential equation requires a Laplace-Beltrami geometric drift \citep{Ikeda1981, Hsu2002}:
\begin{equation}
dx^i = \left( D \sum_j \frac{1}{\sqrt{g}} \frac{\partial}{\partial x^j} \left( \sqrt{g} g^{ij} \right) \right) dt + \sum_j \sigma^i_j dW^j, \label{eq:driftManifold}
\end{equation}
where $x^i$ is the local coordinate of the curved space or manifold 
(for a sphere, these are the polar angle $\theta$ and azimuthal angle $\phi$),
$g_{ij}$ is the inverse metric tensor, $g = \det(g_{ij})$ is the determinant of the metric tensor, and $\sigma^i_j$ is the dispersion matrix, strictly defined by the manifold's geometry through the relation $(\sigma \sigma^T)^{ij} = 2D g^{ij}$.
In our case, the metric determinant $\sqrt{g} = \sin\theta$ generates a purely geometric drift of $D\cot\theta$, reflecting the larger phase-space volume available toward the equator. Substituting this along with the combined torque $F_\theta = F_{\theta,\text{lin}} + F_{\theta,\text{nl}}$ into Equation (12) yields the complete Langevin equation:
\begin{equation}
d\theta = \left[-\Omega(\theta)\sin\theta + D\cot\theta\right] dt + \sqrt{2D} dW_t,
\end{equation}
where $dW_t$ is the Wiener increment and the state-dependent macroscopic relaxation rate emerges explicitly from the competition between the two length scales:
\begin{equation}
\Omega(\theta)=\frac{1}{\gamma_{p}\mu_{0}}\left[\frac{B_{0}^{2}}{L_{\parallel}}-\frac{2B_{0}^{2}}{L_{\delta B}}(1-\cos\theta)\right].
\label{eq:Omega_theta}
\end{equation}
To directly compare this physical derivation with the statistical framework of \citet{Bian2024ApJ...960L...15L}, we map the angular dynamics into the pitch-angle cosine $\mu = \cos\theta$. Applying It\^{o}'s Lemma, $d\mu = -\sin\theta d\theta - \frac{1}{2}\cos\theta (d\theta)^2$, the geometric drift $D\cot\theta$ and the Itô correction $-(1/2)\cos\theta(2D) dt$ exactly combine to yield a noise-induced drift of $-2D\mu$. The Langevin equation in $\mu$-space is therefore:
\begin{equation}
d\mu = \left[ \Omega(\mu)(1-\mu^2) - 2D\mu \right] dt - \sqrt{2D(1-\mu^2)} dW_t.
\label{eq:Langevin-mu}
\end{equation}
Comparing this to the generic drift-diffusion model $d\mu = a_1(\mu) dt + \sqrt{a_2(\mu)} \zeta(t)$ utilized by Bian \& Li (2024), we identify the diffusion coefficient $a_2(\mu) = 2D(1-\mu^2)$ and the drift coefficient $a_1(\mu) = \Omega(\mu)(1-\mu^2) - 2D\mu$. Setting the angular
diffusivity $\nu = 2D$,  the concentration parameter $\kappa = 2\Omega(\mu)/\nu$, 
and using the dimensionless time $\tilde{t} = \nu t$,
equation~(\ref{eq:Langevin-mu}) maps directly to Equation (13) of \citep{Bian2024ApJ...960L...15L}.

While \citet{Bian2024ApJ...960L...15L} assumed a constant $\Omega$ to obtain a single vMF distribution, our derivation here shows that $\Omega(\mu)$ is  $\mu$-dependent.
The steady-state Fokker-Planck solution is $P(\mu) \propto \exp\left( \int (\Omega(\mu')/D) d\mu' \right)$.
Because the physical transition between the linear-dominated and non-linear-balanced regimes is stark, it effectively bifurcates the phase space. Rather than performing a complex continuous integration, approximating this solution as a superposition of two constant-$\Omega$ asymptotic states naturally recovers the observed two-component vMF mixture.

In the small-angle limit ($\mu \to 1$), the non-linear term vanishes and the magnetic tension is dominated by the linear background field operating over $L_\parallel$. This rigid restoring force provides a rapid relaxation rate, $\Omega_{core}$, driving the field direction into the highly concentrated vMF core population ($\kappa_2$). As the wave amplitude grows to large angles ($\mu \to -1$), the non-linear self-tension becomes dynamically significant. Rather than overwhelming the background field and produce an unphysical repulsive force, the spatial scales reorganize (e.g., $L_\parallel$ shrinks to become comparable to $L_{\delta B}$) such that the internal self-tension counter-balances the linear background tension, allowing the structure to approach local force-balance. This cancellation weakens the net restoring torque, collapsing the macroscopic relaxation rate to a much smaller residual value, $\Omega_{tail} < \Omega_{core}$. Consequently the field direction lingers in these large-angle and reversed states, sustaining switchbacks and yielding the broad, weakly confined tail population ($\kappa_1$).

\section{Summary} \label{sec:summary}
In summary, we have shown that the one-point angular distribution of the magnetic field in the highly Alfvénic young solar wind is fundamentally a two-population von Mises-Fisher (vMF) mixture, rather than a single exponential. Using PSP Encounter 1 data, we demonstrated that the distribution consists of a highly concentrated, field-aligned core and a broad, weakly confined tail. The core component is rigorously constrained by the transverse Alfvénic fluctuation level, yielding a concentration parameter $\kappa_2 \simeq 2\langle\vert{}B\vert{}^2\rangle/\langle\delta B_\perp^2\rangle$ that matches observations without any free fitting parameters. Meanwhile, the broad tail component accurately captures the large-amplitude switchback population, confirming that these structures saturate toward $\mu = -1$ in a steady-state drift-diffusion balance.  

The physical origin of these two distinct populations lies in the state-dependent screening of the macroscopic magnetic tension force. By decomposing the restoring torque into a linear background term and a non-linear self-tension term, we revealed a dynamic competition governed by two distinct correlation length scales. At small deflection angles, the rigid linear tension operating over the parallel scale $L_\parallel$ provides a strong, constant restoring rate, dynamically confining the core population. At large angles, however, the structure approaches local force-balance as the non-linear self-tension, acting over the active perturbation scale $L_{\delta B}$, grows to directly counter-balance the background tension. This internal cancellation effectively collapses the macroscopic relaxation rate, allowing the field direction to linger in reversed states and naturally sustaining the long-lived switchback tail. This physics-based two-population framework connects macroscopic fluid parameters ($L_{||}$ and $L_{\delta B}$) directly to those describing the directional statistics ($\kappa_1$ and $\kappa_2$), offering a powerful diagnostic for probing the structural evolution of MHD turbulence and switchbacks in the inner heliosphere.

\textit{Acknowledgments.}---This work is supported, at MUST, in part by FDCT grants 0022/2025/ASJ, 0137/2025/AFJ, and FRG-25-113-SSI, and by the S\~ao Paulo Research Foundation (FAPESP), {Process Number 2025/11756-6}. RAM acknowledges support from CNPq (407341/2022-6, 407493/2022-0), FAPDF (383/2023) and Brazilian Space Agency (TED 000529/2024). ELR acknowledges support from CNPq (306265/2024-9). B.R. was supported by FAPESP under grants 2023/16513-9 and 2024/20546-2.


\bibliographystyle{aasjournal}
\bibliography{Population}

@PREAMBLE{
 "\providecommand{\noopsort}[1]{}" 
 # "\providecommand{\singleletter}[1]{#1}%" 

}

@book{Hsu2002,
  author    = {Hsu, Elton P.},
  title     = {Stochastic Analysis on Manifolds},
  series    = {Graduate Studies in Mathematics},
  volume    = {38},
  publisher = {American Mathematical Society},
  year      = {2002},
  address   = {Providence, RI}
}

@book{Ikeda1981,
  author    = {Ikeda, Nobuyuki and Watanabe, Shinzo},
  title     = {Stochastic Differential Equations and Diffusion Processes},
  series    = {North-Holland Mathematical Library},
  volume    = {24},
  publisher = {North-Holland},
  year      = {1981},
  address   = {Amsterdam}
}

@ARTICLE{Li2023ApJ...945...150B,
       author = {{Li}, Gang and {Bian}, N.~H.},
        title = "{Lagrangian Stochastic Model for the Motions of Magnetic Footpoints on the Solar Wind Source Surface and the Path Lengths of Boundary-driven Interplanetary Magnetic Field Lines}",
      journal = {The Astrophysical Journal},
         year = 2023,
        month = mar,
       volume = {945},
       number = {2},
          eid = {150},
        pages = {150},
          doi = {10.3847/1538-4357/acbd43},
       adsurl = {https://ui.adsabs.harvard.edu/abs/2023ApJ...945..150L}
}

@ARTICLE{Xu2015ApJ...801...58L,
       author = {{Xu}, F. and {Li}, G. and {Zhao}, L. and {Zhang}, Y. and {Khabarova}, O. and {Miao}, B. and {le Roux}, J.},
        title = "{Angular Distribution of Solar Wind Magnetic Field Vector at 1 AU}",
      journal = {The Astrophysical Journal},
         year = 2015,
        month = mar,
       volume = {801},
       number = {1},
          eid = {58},
        pages = {58},
          doi = {10.1088/0004-637X/801/1/58},
       adsurl = {https://ui.adsabs.harvard.edu/abs/2015ApJ...801...58X}
}

@ARTICLE{Chang2022ApJ...931...105H,
       author = {{Chang}, Qing and {Xu}, Xiaojun and {Wang}, Xing and {Ye}, Yudong and {Xu}, Qi and {Wang}, Jing and {Wang}, Ming and {Zhou}, Zilu and {Luo}, Lei and {Cheng}, Shaoguan and {He}, Peishan},
        title = "{The Solar Wind Parker Spiral Angle Distributions and Variations at 1 au}",
      journal = {The Astrophysical Journal},
         year = 2022,
        month = jun,
       volume = {931},
       number = {2},
          eid = {105},
        pages = {105},
          doi = {10.3847/1538-4357/ac6bf3},
       adsurl = {https://ui.adsabs.harvard.edu/abs/2022ApJ...931..105C}
}

@ARTICLE{Borovsky2010JGR...115...A09101B,
       author = {{Borovsky}, Joseph E.},
        title = "{On the variations of the solar wind magnetic field about the Parker spiral direction}",
      journal = {Journal of Geophysical Research (Space Physics)},
         year = 2010,
        month = sep,
       volume = {115},
       number = {A9},
          eid = {A09101},
        pages = {A09101},
          doi = {10.1029/2009JA015040},
       adsurl = {https://ui.adsabs.harvard.edu/abs/2010JGRA..115.9101B}
}

@ARTICLE{Belcher1971JGR...76...3534D,
       author = {{Belcher}, J.~W. and {Davis}, Jr., Leverett},
        title = "{Large-amplitude Alfv{\'e}n waves in the interplanetary medium, 2}",
      journal = {Journal of Geophysical Research},
         year = 1971,
        month = jan,
       volume = {76},
       number = {16},
        pages = {3534},
          doi = {10.1029/JA076i016p03534},
       adsurl = {https://ui.adsabs.harvard.edu/abs/1971JGR....76.3534B}
}

@ARTICLE{Barnes1981JGR...86...7498B,
       author = {{Barnes}, A.},
        title = "{Interplanetary Alfv{\'e}nic fluctuations: A stochastic model}",
      journal = {Journal of Geophysical Research},
         year = 1981,
        month = sep,
       volume = {86},
       number = {A9},
        pages = {7498-7506},
          doi = {10.1029/JA086iA09p07498},
       adsurl = {https://ui.adsabs.harvard.edu/abs/1981JGR....86.7498B}
}

@ARTICLE{Bale2019Nat...576...237W,
       author = {{Bale}, S.~D. and {Badman}, S.~T. and {Bonnell}, J.~W. and {Bowen}, T.~A. and {Burgess}, D. and {Case}, A.~W. and {Cattell}, C.~A. and {Chandran}, B.~D.~G. and {Chaston}, C.~C. and {Chen}, C.~H.~K. and {Drake}, J.~F. and {de Wit}, T. Dudok and {Eastwood}, J.~P. and {Ergun}, R.~E. and {Farrell}, W.~M. and {Fong}, C. and {Goetz}, K. and {Goldstein}, M. and {Goodrich}, K.~A. and {Harvey}, P.~R. and {Horbury}, T.~S. and {Howes}, G.~G. and {Kasper}, J.~C. and {Kellogg}, P.~J. and {Klimchuk}, J.~A. and {Korreck}, K.~E. and {Krasnoselskikh}, V.~V. and {Krucker}, S. and {Laker}, R. and {Larson}, D.~E. and {MacDowall}, R.~J. and {Maksimovic}, M. and {Malaspina}, D.~M. and {Martinez-Oliveros}, J. and {McComas}, D.~J. and {Meyer-Vernet}, N. and {Moncuquet}, M. and {Mozer}, F.~S. and {Phan}, T.~D. and {Pulupa}, M. and {Raouafi}, N.~E. and {Salem}, C. and {Stansby}, D. and {Stevens}, M. and {Szabo}, A. and {Velli}, M. and {Woolley}, T. and {Wygant}, J.~R.},
        title = "{Highly structured slow solar wind emerging from an equatorial coronal hole}",
      journal = {Nature},
         year = 2019,
        month = dec,
       volume = {576},
       number = {7786},
        pages = {237-242},
          doi = {10.1038/s41586-019-1818-7},
       adsurl = {https://ui.adsabs.harvard.edu/abs/2019Natur.576..237B}
}

@ARTICLE{DudokdeWit2020ApJS...246...39W,
       author = {{Dudok de Wit}, Thierry and {Krasnoselskikh}, Vladimir V. and {Bale}, Stuart D. and {Bonnell}, John W. and {Bowen}, Trevor A. and {Chen}, Christopher H.~K. and {Froment}, Clara and {Goetz}, Keith and {Harvey}, Peter R. and {Jagarlamudi}, Vamsee Krishna and {Larosa}, Andrea and {MacDowall}, Robert J. and {Malaspina}, David M. and {Matthaeus}, William H. and {Pulupa}, Marc and {Velli}, Marco and {Whittlesey}, Phyllis L.},
        title = "{Switchbacks in the Near-Sun Magnetic Field: Long Memory and Impact on the Turbulence Cascade}",
      journal = {The Astrophysical Journal Supplement Series},
         year = 2020,
        month = feb,
       volume = {246},
       number = {2},
          eid = {39},
        pages = {39},
          doi = {10.3847/1538-4365/ab5853},
archivePrefix = {arXiv},
       eprint = {1912.02856},
 primaryClass = {astro-ph.SR},
       adsurl = {https://ui.adsabs.harvard.edu/abs/2020ApJS..246...39D}
}

@ARTICLE{Fargette2021ApJ...919...96B,
       author = {{Fargette}, Na{\"\i}s and {Lavraud}, Benoit and {Rouillard}, Alexis P. and {R{\'e}ville}, Victor and {Dudok De Wit}, Thierry and {Froment}, Clara and {Halekas}, Jasper S. and {Phan}, Tai D. and {Malaspina}, David M. and {Bale}, Stuart D. and {Kasper}, Justin C. and {Louarn}, Philippe and {Case}, Anthony W. and {Korreck}, Kelly E. and {Larson}, Davin E. and {Pulupa}, Marc and {Stevens}, Michael L. and {Whittlesey}, Phyllis L. and {Berthomier}, Matthieu},
        title = "{Characteristic Scales of Magnetic Switchback Patches Near the Sun and Their Possible Association With Solar Supergranulation and Granulation}",
      journal = {The Astrophysical Journal},
         year = 2021,
        month = oct,
       volume = {919},
       number = {2},
          eid = {96},
        pages = {96},
          doi = {10.3847/1538-4357/ac1112},
archivePrefix = {arXiv},
       eprint = {2109.01519},
 primaryClass = {astro-ph.SR},
       adsurl = {https://ui.adsabs.harvard.edu/abs/2021ApJ...919...96F}
}

@ARTICLE{Fargette2022A&A...663...A109K,
       author = {{Fargette}, Na{\"\i}s and {Lavraud}, Benoit and {Rouillard}, Alexis P. and {R{\'e}ville}, Victor and {Bale}, Stuart D. and {Kasper}, Justin},
        title = "{The preferential orientation of magnetic switchbacks and its implications for solar magnetic flux transport}",
      journal = {Astronomy and Astrophysics},
         year = 2022,
        month = jul,
       volume = {663},
          eid = {A109},
        pages = {A109},
          doi = {10.1051/0004-6361/202243537},
archivePrefix = {arXiv},
       eprint = {2203.14591},
 primaryClass = {astro-ph.SR},
       adsurl = {https://ui.adsabs.harvard.edu/abs/2022A&A...663A.109F}
}

@ARTICLE{Bian2024ApJ...960L...15L,
       author = {{Bian}, N.~H. and {Li}, Gang},
        title = "{A Fokker-Planck Framework for Studying the Variability of the Magnetic Field Direction in the Alfv{\'e}nic Streams of the Solar Wind}",
      journal = {The Astrophysical Journal Letters},
         year = 2024,
        month = jan,
       volume = {960},
       number = {2},
          eid = {L15},
        pages = {L15},
          doi = {10.3847/2041-8213/ad17cc},
       adsurl = {https://ui.adsabs.harvard.edu/abs/2024ApJ...960L..15B}
}

@article{Borovsky2008JGR...13...A08110B,
  author    = {Borovsky, Joseph E.},
  title     = {Flux tube texture of the solar wind: Strands of the magnetic carpet at 1 AU?},
  journal   = {Journal of Geophysical Research: Space Physics},
  volume    = {13},
  number    = {A8},
  pages     = {A08110},
  year      = {2008},
  month     = {August},
  doi       = {10.1029/2007JA012684},
  url       = {https://agupubs.onlinelibrary.wiley.com/doi/abs/10.1029/2007ja012684}
}

@INPROCEEDINGS{Bavassano1983,
       author = {{Bavassano}, B. and {Mariani}, F.},
        title = "{Interplanetary Alfv{\'e}nic fluctuations: a statistical study of the directional variations of the magnetic field.}",
    booktitle = {NASA Conference Publication},
         year = 1983,
       editor = {{Neugebauer}, M.},
       series = {NASA Conference Publication},
       volume = {228},
        month = nov,
        pages = {0..99},
       adsurl = {https://ui.adsabs.harvard.edu/abs/1983NASCP.2280..99B}
}

@ARTICLE{Goldreich1995ApJ...438...763S,
       author = {{Goldreich}, P. and {Sridhar}, S.},
        title = "{Toward a Theory of Interstellar Turbulence. II. Strong Alfvenic Turbulence}",
      journal = {The Astrophysical Journal},
         year = 1995,
        month = jan,
       volume = {438},
        pages = {763},
          doi = {10.1086/175121},
       adsurl = {https://ui.adsabs.harvard.edu/abs/1995ApJ...438..763G}
}

@ARTICLE{Parker1958ApJ...128...664P,
       author = {{Parker}, E.~N.},
      journal = {The Astrophysical Journal},
         year = 1958,
        month = nov,
       volume = {128},
        pages = {664},
          doi = {10.1086/146579},
       adsurl = {https://ui.adsabs.harvard.edu/abs/1958ApJ...128..664P}
}

@ARTICLE{Jokipii1966ApJ...146...480J,
       author = {{Jokipii}, J.~R.},
      journal = {The Astrophysical Journal},
         year = 1966,
        month = nov,
       volume = {146},
        pages = {480},
          doi = {10.1086/148912},
       adsurl = {https://ui.adsabs.harvard.edu/abs/1966ApJ...146..480J}
}

@ARTICLE{Bale2016SSR...204...49W,
       author = {{Bale}, S.~D. and {Goetz}, K. and {Harvey}, P.~R. and {Turin}, P. and {Bonnell}, J.~W. and {Dudok de Wit}, T. and {Ergun}, R.~E. and {MacDowall}, R.~J. and {Pulupa}, M. and {Andre}, M. and {Bolton}, M. and {Bougeret}, J.-L. and {Bowen}, T.~A. and {Burgess}, D. and {Cattell}, C.~A. and {Chandran}, B.~D.~G. and {Chaston}, C.~C. and {Chen}, C.~H.~K. and {Choi}, M.~K. and {Connerney}, J.~E.~P. and {Cranmer}, S. and {Diaz-Aguado}, M. and {Drake}, J.~F. and {Farrell}, W.~M. and {Fergeau}, P. and {Fermin}, J. and {Fischer}, J. and {Fox}, N. and {Goodrich}, K.~A. and {Halekas}, J.~S. and {Horbury}, T.~S. and {Howes}, G.~G. and {Jagarlamudi}, V.~K. and {Kasper}, J.~C. and {Kellogg}, P.~J. and {Korreck}, K.~E. and {Krasnoselskikh}, V.~V. and {Larson}, D.~E. and {Livi}, R. and {Ludlam}, M. and {Malaspina}, D.~M. and {Maruca}, B.~A. and {McManus}, M.~D. and {Meyer-Vernet}, N. and {Moncuquet}, M. and {Mozer}, F.~S. and {Raouafi}, N.~E. and {Salem}, C. and {Schwadron}, N. and {Stevens}, M. and {Szabo}, A. and {Velli}, M. and {Woolley}, T. and {Wygant}, J.~R.},
        title = "{The FIELDS Instrument Suite for Solar Probe Plus: Measuring the Coronal Plasma and Magnetic Field, Plasma Waves and Turbulence, and Radio Signatures of Solar Transients}",
      journal = {Space Science Reviews},
         year = 2016,
        month = dec,
       volume = {204},
       number = {1-4},
        pages = {49-82},
          doi = {10.1007/s11214-016-0244-5},
       adsurl = {https://ui.adsabs.harvard.edu/abs/2016SSRv..204...49B}
}

@ARTICLE{Ness1967SoPh....2..351N,
       author = {{Ness}, Norman F. and {Wilcox}, John M.},
        title = "{Interplanetary sector structure, 1962 1966}",
      journal = {Solar Physics},
         year = 1967,
        month = nov,
       volume = {2},
       number = {3},
        pages = {351-359},
          doi = {10.1007/BF00147849},
       adsurl = {https://ui.adsabs.harvard.edu/abs/1967SoPh....2..351N}
}

@ARTICLE{Ness1966ApJ...143...23N,
       author = {{Ness}, Norman F. and {Wilcox}, John M.},
        title = "{Extension of the Photospheric Magnetic Field Into Interplanetary Space}",
      journal = {The Astrophysical Journal},
         year = 1966,
        month = jan,
       volume = {143},
        pages = {23},
          doi = {10.1086/148473},
       adsurl = {https://ui.adsabs.harvard.edu/abs/1966ApJ...143...23N}
}

@article{Barnes1976,
    author = {Barnes, A.},
    title = {On the nonexistence of plane-polarized large amplitude Alfv{\'e}n waves},
    journal = {Journal of Geophysical Research},
    volume = {81},
    number = {1},
    pages = {281--282},
    year = {1976},
    doi = {10.1029/JA081i001p00281}
}

@article{Baker1984,
    author = {Baker, S. and Cousins, R. D.},
    title = {Clarification of the use of chi-square and likelihood functions in fits to histograms},
    journal = {Nuclear Instruments and Methods in Physics Research},
    volume = {221},
    number = {2},
    pages = {437--442},
    year = {1984},
    doi = {10.1016/0167-5087(84)90016-4}
}

\end{document}